# Frequency stability and phase noise measurements of a 5,860 km-long intercontinental seafloor optical fibre cable

Giuseppe Marra[1], Paul Gaynor[1], Mattia Cantono[2], Valey Kamalov[2], Sean Mulholland[1], Ian Hill[1], Marco Schioppo[1], Jacques-Olivier Gaudron[1], Irene-Barbeito Edreira[1], Cecilia Clivati[3], Davide Calonico[3]

[1]National Physical Laboratory, Teddington, TW11 0LW, United Kingdom
[2]Google LLC, Mountain View, CA, USA.
[3]Istituto Nazionale di Ricerca Metrologica, Strada delle Cacce 91, 10135 Torino, Italy

Corresponding author: giuseppe.marra@npl.co.uk

**Optical clock comparison via optical fibre links has been achieved over continental scales, but has not yet been demonstrated intercontinentally. The transfer of ultra-stable optical frequencies over transoceanic distances is a challenging task, as the seafloor cable architecture prevents the use of environmental noise suppression techniques currently employed on land-based metrological links. As a fundamental first step towards devising suitable frequency transfer techniques to enable future clock comparison on a global scale, here we show the free-running frequency stability and phase noise measurements of a transatlantic seafloor optical link between the UK and Canada. To the best of our knowledge, these are the first ever measurements of an intercontinental optical fibre link.**

Today, the comparison of optical clocks using optical fibre links can be routinely performed over continental scales [1-5]. However, optical fibre-based clock comparison over intercontinental scales is yet to be demonstrated and cannot be achieved with the transfer techniques currently employed on fibre links on land. In fact, these techniques require bidirectional propagation of the light in the same fibre to achieve sufficient suppression of environmentally-induced noise in the fibre in order to meet the demanding performance suitable for optical clock comparison [6]. On land, bidirectional propagation of light in the same fibre can be achieved with reasonable effort. In this case, the standard unidirectional telecommunication optical amplifiers installed along telecommunication links can be replaced with bidirectional ones. However, this is not a viable solution for the optical amplifiers installed along seafloor cables and different frequency transfer techniques need to be used.

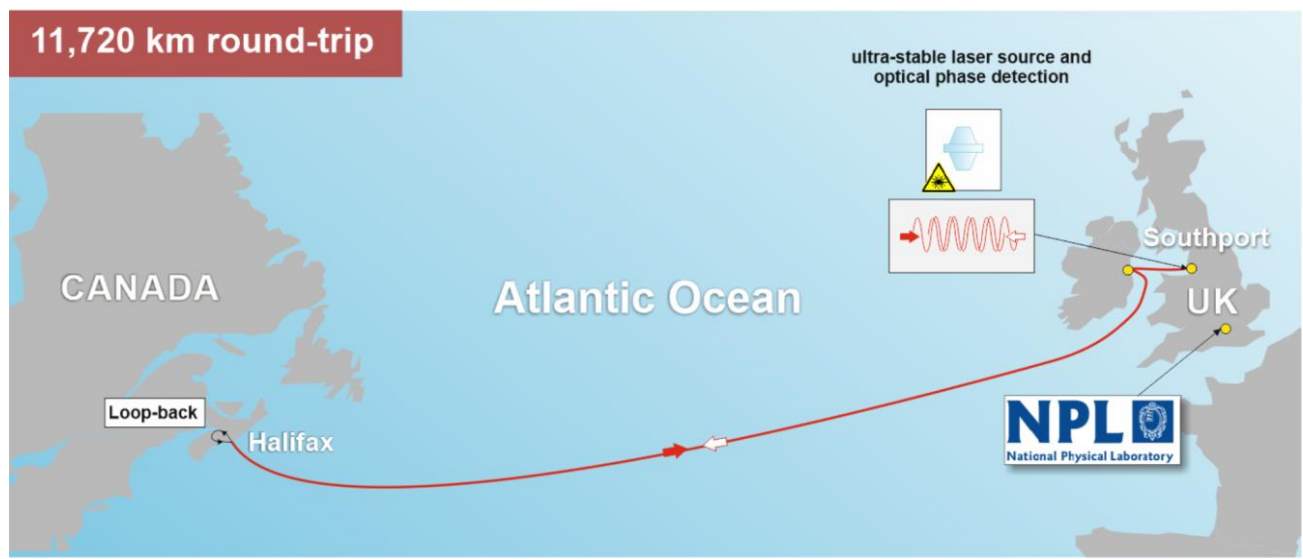

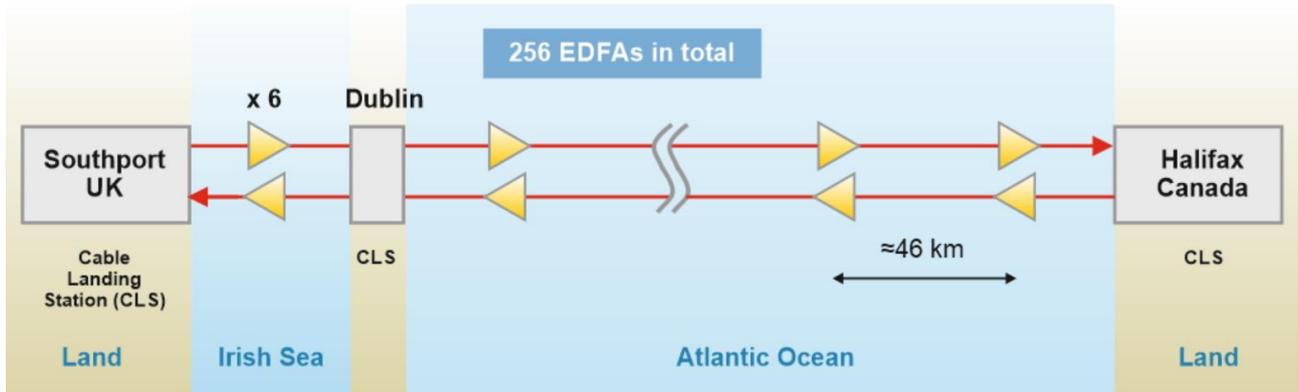

**Fig. 1. A**) Illustration of the intercontinental optical fibre cable between UK and Canada. **B**) Architecture of the intercontinental link, constituting of a fibre pair. The yellow triangles represent optical amplifiers. The average span length is 46 km. CLS: Cable Landing Station.

Whilst the level of environmentally-induced frequency stability and phase noise data is now available for a number of optical links installed on land, only short seafloor links (~100 km) have been characterized so far [7]. Knowledge of the level of noise that can be found on an optical link spanning intercontinental distances is a crucial first step to devise suitable noise suppression techniques and estimate what level of frequency transfer performance could be achieved with them. However, to the best of our knowledge, these crucial measurements have not been performed until now.

In this work we present the free running frequency stability and phase noise of a 5,860 km-long transatlantic optical fibre cable between the UK and Canada (Fig. 1A). The optical fibre link consists of a fibre pair, one fibre for each direction of propagation, within a multi-pair optical cable. The seafloor cable is organised in two sections: a 248 km-long cable from cable landing stations (CLS) in Southport, UK, and Dublin, Ireland, and a 5,612 km-long cable from Dublin to Halifax, Canada. A total of 128 erbium-doped fibre amplifier (EDFA)-based repeaters are installed on the link to compensate for the fibre loss, 6 in the Southport-Dublin and 122 in the Dublin-Halifax sections (Fig. 1B). The average span between repeaters is 46 km. Land-installed sections of the cable at both the Canada and UK end cumulatively account for less than 33 km. A beach manhole joins the submarine cable with the terrestrial cable which runs in underground ducts to the CLS. The optical link consists of a combination of Large Mode Field (LMF), High Dispersion (HDF) and Non-dispersion shifted (NDSF) fibres, for dispersion and non-linearity management of the link.

The cable runs for approximately 1,000 km and 250 km on the European and North American continental shelves respectively. The remaining ~4,600 km of the cable runs in deep water with a depth of up to 5,200 m. The cable crosses the Mid-Atlantic Ridge where the depth of the seafloor varies from 3,740 to 2,330 m. The cable employs different levels of armour depending on water depth, ranging from heavy rock-armour near the coastlines, where damage from ship anchors and fishing trawlers is more likely, to lightweight armour in deeper waters, where the cable diameter is less than 20 mm. To the best of our knowledge, in shallow waters the cable was buried for the majority of its length at a depth ranging from 0.3 to 1.8 m, carried out during its installation in early 2000s. In deep waters (depth greater than 1,500 m) it is simply resting on the sea floor. Over the years, it is conceivable that deep water sections might become covered by a layer of soft sediments transported by currents.

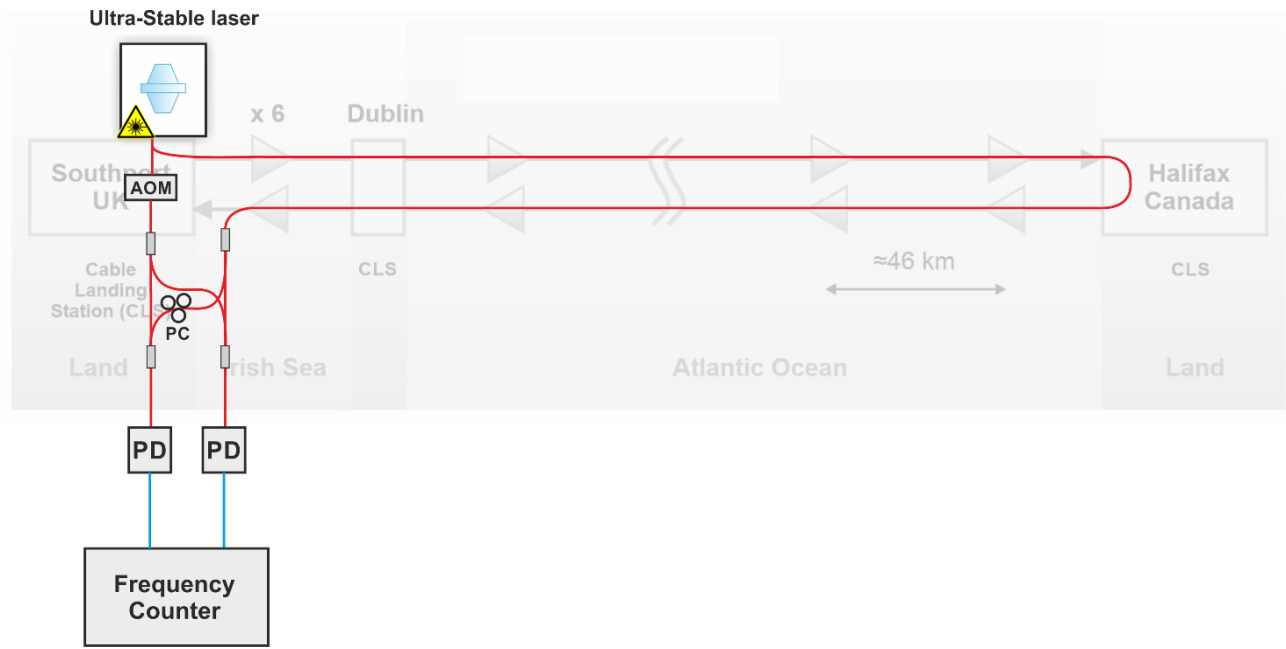

**Fig. 2:** Interferometric setup used to measure the frequency stability and phase noise of the intercontinental optical fibre cable/ PD: Photodetector, AOM: Acousto-Optic Modulator, PC: Polarisation controller

The interferometric setup used for the phase noise and frequency stability measurement is shown in Fig. 2 and was installed in the cable landing station in Southport in October 2020. The ultra-stable optical frequency source, designed and built at NPL, consists of a commercial narrow-linewidth fibre laser phase locked to a high-finesse transportable Fabry-Perot Ultra-Low Expansion (ULE) glass cavity [8]. The fractional frequency instability of the phase-locked laser system is below $5 \times 10^{-15}$ at 1 s and the residual drift is below 100 mHz/s.

The ultra-stable laser light at 1543.33 nm generated by the optical reference was injected into the corresponding 50 GHz-wide Dense Wavelength-Division Multiplexing (DWDM) channel of the optical fibre link with a power less than a few dBm. Live data traffic was present on several of the other channels, including the neighbouring ones. No bit-error-rate degradation were reported on the live channels. At the Halifax end, an optical patch joining the fibre pair allows for the outgoing light to be re-injected in the fibre link in the opposite direction of propagation, forming a 11,720 km-long loop from the Southport end. Here, the incoming light is combined with that from the ULE cavity-based optical reference on a photodetector so that environmentally-induced optical phase perturbations accumulated along the link can be measured. In order to create a radio-frequency (RF) beat that can be directly measured by a phase/frequency counter, the local reference is frequency shifted by 40 MHz with an acousto-optic modulator. The resulting RF beat after a round trip is then filtered with a few MHz wide band pass filter and measured by a dead-time free RF frequency counter.

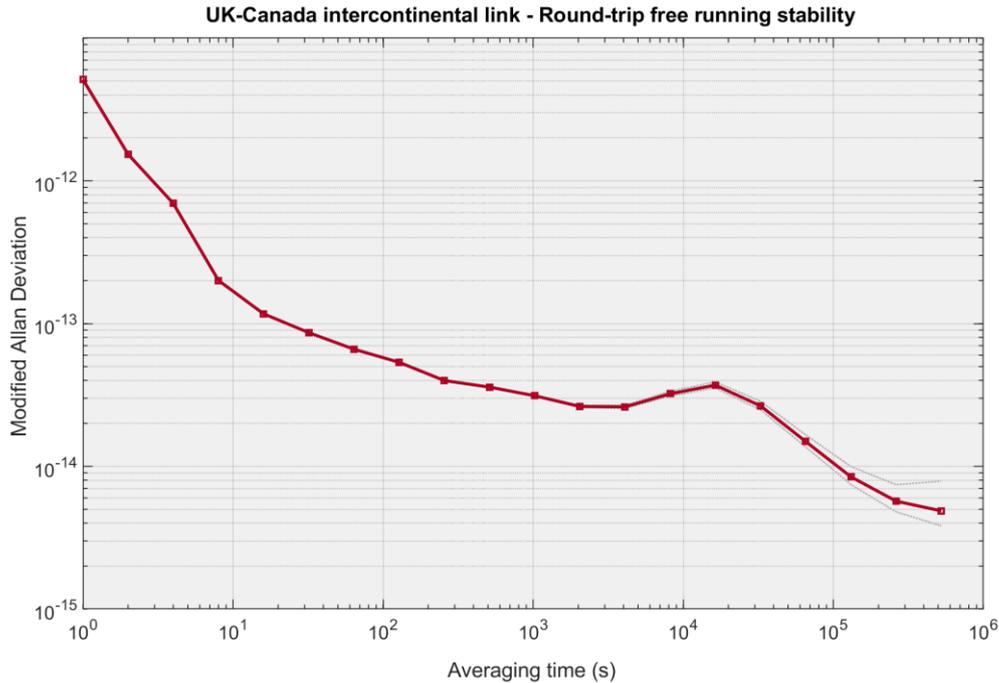

**Fig.3:** Environmentally-induced frequency instability of the UK-Canada link

Slow polarization changes induced by temperature changes of the seafloor and land sections of the cable cause the SNR of the RF beat to degrade over time. However, we note that the SNR is sufficiently high to be correctly counted for a substantial fraction of a day, as also previously reported on shorter links [7, 9]. To enable continuous long-term measurements, the optical signal returned after a round trip in the link is combined with the local reference at two different polarizations, obtaining two separate RF beats. In post-processing, at any point in time, the most suitable beat is selected when the SNR of the other beat is insufficient to correctly drive the frequency counter. The polarization difference is manually set such that the SNR on one beat is minimised when that on the other beat is maximised. The fractional frequency stability of the 11,720 km-long UK-Canada round-trip intercontinental link, measured over the 2 weeks, is shown in Fig. 3. Despite the length of the seafloor link being several times that of available metrological links on land, the long-term frequency stability is measured to be of comparable or even higher level [10-14]. We attribute the exceptional longer-term stability to the high temperature stability of the seafloor, which can vary by as little as 300 mK over time scales of several years in deep water [15-17]. Shallow coastal waters will exhibit higher temperature variability.

The phase noise of the intercontinental link is shown in Fig. 4. The short-term stability of the intercontinental link was characterised by phase noise measurements obtained from the phase/frequency counter with a 1 ms gate time. For offset frequencies from 1 mHz to 0.1 Hz the phase noise exhibits a $1/f^2$ slope, indicating a random walk of phase behaviour. We observe a substantial amount of noise in the 0.1 Hz to 3 Hz region, with a number of narrow bandwidth features. For frequencies higher than 3 Hz the noise level drops significantly and no longer exhibits narrow-bandwidth features.

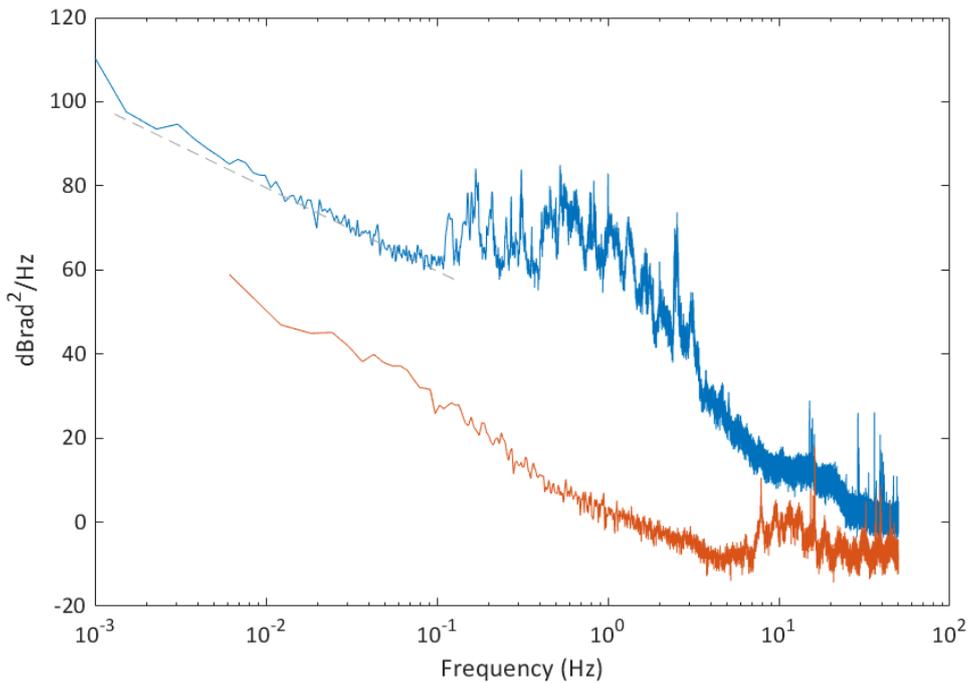

**Fig.4:** Environmentally-induced phase noise of the UK-Canada intercontinental seafloor link (blue). Cable landing station monitoring loop for comparison (red).

The CLS exhibits a high level of acoustic noise, arising from air conditioning and equipment fans. In order to measure the contribution of this acoustic noise to the measured phase noise and stability of the UK-Canada link, a 50 m-long optical fibre pair was installed between the equipment rack used for the experiments and the location at which the seafloor cable enters the CLS. This additional fibre loop follows the same route as the fibre pair used to link the equipment to the seafloor cable. The environmentally-induced phase variations on this loop, which also include temperature and pressure changes in the CLS, are measured synchronously and subtracted from those on the intercontinental link loop. As expected, the noise measured on the CLS optical loop is substantially below the noise of the UK-Canada link for offset frequencies up to a several Hz. For higher frequencies, some contribution of the noise induced on the link in cable landing stations cannot be safely excluded. Indeed, additional noise from the cable landing stations in Halifax and Dublin might increase the noise showing in Fig.4 in this frequency range. A similar level of noise can be expected in Halifax with respect to Southport. However, a higher level could be expected from the cable landing station in Dublin as it is a larger and busier facility.

**Conclusions**

We have, to the best of our knowledge, performed the first measurement of the phase noise and frequency stability of an intercontinental fibre link. Despite the round-trip optical path exceeding 11,000 km, a very high long-term fractional frequency stability better than $1\times10^{-14}$ was measured at time scales exceeding one day, which we attribute to the very high stability of the deep-water seafloor temperature. We believe that these measurements constitute the basis to develop suitable frequency transfer techniques for optical frequency comparison over intercontinental scales using seafloor cables, towards future global timescales based on optical clocks.

**Funding**

NPL: This work was supported by an ISCF Metrology Fellowship grant and as part of the National Measurement System Programme provided by the UK government's Department for Business, Energy and Industrial Strategy (BEIS), and by an Award from the National Physical Laboratory's Directors' Science and Engineering Fund. INRIM: This project receives funding from the European Union's Horizon 2020 research and innovation programme under Grant Agreement No. 951886 (CLONETS-DS).